\documentclass{article}
\usepackage{spconf,amsmath,graphicx}
\usepackage{url}
\usepackage{amssymb}
\usepackage{multirow}
\usepackage{caption}
\usepackage{subcaption}
\usepackage{color}
\usepackage{tikz}
\usepackage{booktabs}
\usepackage{enumitem}
\usepackage{multirow}
\usepackage{hyperref}
\usepackage{multicol}
\usepackage{pifont}

\title{TONet: Tone-Octave Network for Singing Melody Extraction \\ from Polyphonic Music}
\name{Ke Chen $^1$, Shuai Yu $^2$, Cheng-i Wang $^3$, Wei Li $^2$, Taylor Berg-Kirkpatrick $^1$, Shlomo Dubnov $^1$}
\address{
         $^1$University of California San Diego \quad
	     $^2$Fudan University \quad
	     $^3$Smule Inc.
	     }

\begin{document}
%
\maketitle

\begin{abstract}
Singing melody extraction is an important problem in the field of music information retrieval. Existing methods typically rely on frequency-domain representations to estimate the sung frequencies. However, this design does not lead to human-level performance in the perception of melody information for both tone (pitch-class) and octave. In this paper, we propose TONet\footnote{\href{https://github.com/RetroCirce/TONet}{code available at https://github.com/RetroCirce/TONet}}, a plug-and-play model that improves both tone and octave perceptions by leveraging a novel input representation and a novel network architecture. First, we present an improved input representation, the Tone-CFP, that explicitly groups harmonics via a rearrangement of frequency-bins. Second, we introduce an encoder-decoder architecture that is designed to obtain a salience feature map, a tone feature map, and an octave feature map. Third, we propose a tone-octave fusion mechanism to improve the final salience feature map. Experiments are done to verify the capability of TONet with various baseline backbone models. Our results show that tone-octave fusion with Tone-CFP can significantly improve the singing voice extraction performance across various datasets -- with substantial gains in octave and tone accuracy. 
\end{abstract}

\vspace{-0.2cm}
\section{Introduction}
\vspace{-0.2cm}
\label{sec:intro}

Singing melody extraction is an important task in the music information retrieval field.  The task is to estimate the fundamental frequency (F0) of the leading singing voice from an audio waveform. Melody extraction has many downstream applications, including music recommendation \cite{ke-recom}, cover song identification \cite{ke-cover2}, and music generation \cite{ke-thesis, ke-sketchnet}. The presence of background sounds and musical accompaniments of different instruments are interwoven with the leading voice, making this task quite challenging \cite{TabrikianDD04}. And one of the challenges is the lacking of representations that isolate the singing melody efficiently.

In order to find such representations, previous works \cite{kum2016melody,lu2018vocal,su2018vocal,chen2019cnn},  have leveraged the power of deep neural networks, mainly the convolutional neural network (CNN). Also, musical and structural prior knowledge are used to improve the performance. For example, voiced and unvoiced frames can be independently recognized \cite{hsieh2018streamlined} or jointly detected via classification tasks \cite{jdc-melody}. The relationship between frequencies can be further captured through multi-dilation \cite{GaoYC20}, attention networks \cite{ftanet}, or harmonic constant-Q transform (HCQT) \cite{bittner2017deep}. These models no longer rely solely on direct supervision -- they also leverage inductive biases in model design.

We focus on two aspects of musical prior knowledge to improve melody extraction. First, current methods are prone to incorrectly predict the \textbf{tone} within the correct octave. We refer to this type of error as a \textbf{tone error}\footnote{In this paper, we define the term \textbf{tone} as pitch-class (C,C\#,D,...,B)} (e.g., mistake \textit{A4} as \textit{G4} or \textit{A\#4}, etc.). 
Second, existing methods suffer from \textbf{octave errors} -- even in cases where the sung tone is correctly predicted (e.g., mistake \textit{A4} as \textit{A3} or \textit{A5}). When the predicted pitch contours are visualized, octave errors are manifest as large jumps in the pitch contours. Based on human auditory research \cite{tone-octave-percp}, a human can separately perceive a musical note in both tone and octave aspects. This motivates us to explore models that factorize the task into tone and octave prediction problems. Musically, a tone is determined by the harmonic position and distance. An octave is determined by the frequency band where the fundamental frequency (the 1$^{st}$ harmonic) lies. This motivates us to design a representation that allows the networks to capture critical information for tone and octave predictions more directly. 

In this paper, we propose the TONet with three contributions. First, based on the combined frequency and periodicity representation (CFP) \cite{su2018vocal, hsieh2018streamlined}, we propose a new representation, Tone-CFP (TCFP), which rearranges the tonal harmonics into adjacent bins, thus allows a CNN layer to capture harmonic relationships in its receptive field. Secondly, we propose a new scheme that factorizes pitch prediction into separate tone and octave pathways and losses in our neural network architecture. By leveraging self-attention modules \cite{vaswani2017attention}, we use a salience feature map extracted from CFP and TCFP to predict the tone and the octave, respectively. The final frequency-bin predictions and loss are formed by fusing tone, octave, and salience feature maps. Thirdly, we conduct a comprehensive experiment to show that many existing backbones can be inserted to our plug-and-play TONet to drive melody extraction.


\section{Proposed model}
\vspace{-0.2cm}
\subsection{Tone-CFP Representation}
We follow \cite{hsieh2018streamlined} to employ the CFP representation. It contains three parts: a power-scaled spectrogram, a generalized cepstrum (GC) and a generalized cepstrum of spectrum (GCoS). Due to the local receptive field of CNN, most existing methods \cite{lu2018vocal,su2018vocal,chen2019cnn,hsieh2018streamlined} cannot capture harmonic relationships well. Musically, harmonics play a huge role in tone perception since the pitches that hold the same tone share similar harmonics. To solve this issue, we extract Tone-CFP by gathering together the CFP frequency bins that belong to the same tone. Formally, for a CFP representation $spec$, we assume that it has $L$ frequency bins per octave and $F$ bins in total. For the $i$-th frequency bin $spec_i$ in $spec$, it belongs to the $k$-th \textbf{non-overlapping} set $S_k$ by the following definition:
\begin{equation}
    S_k = \{spec_i \, |\, i \equiv k \;\mathrm{mod}\; L \}, k = 0,1,2,..., L-1 
    \vspace{-0.2cm}
\end{equation}

After we group all $spec_i$ into different sets, we sort all sets $S$ according to their indexes to obtain the TCFP representation. Logically, TCFP is constructed by rearranging the order of CFP on the frequency axis. Note that this approach can be also applied to other spectrum features (e.g. STFT, CQT, etc.). With this rearrangement, the frequency bins corresponding to the same tone's harmonics are grouped in the same region. Thus they can be captured by the local receptive field of 2D-CNN kernels. As shown in the top of Figure \ref{fig:model_arch}, TCFP has a rough tone (pitch-class) contour in its first layer. Compared with the Chroma feature \cite{chromafeature}, TCFP has three channels representing different detailed information. Also, it can be seamlessly integrated into the model because it has the same size as the original representation (i.e. CFP).

\vspace{-0.5cm}
\subsection{The TONet Encoder}
\begin{figure}[t]
	\centering
	\vspace{-0.7cm}
	\includegraphics[width=\columnwidth]{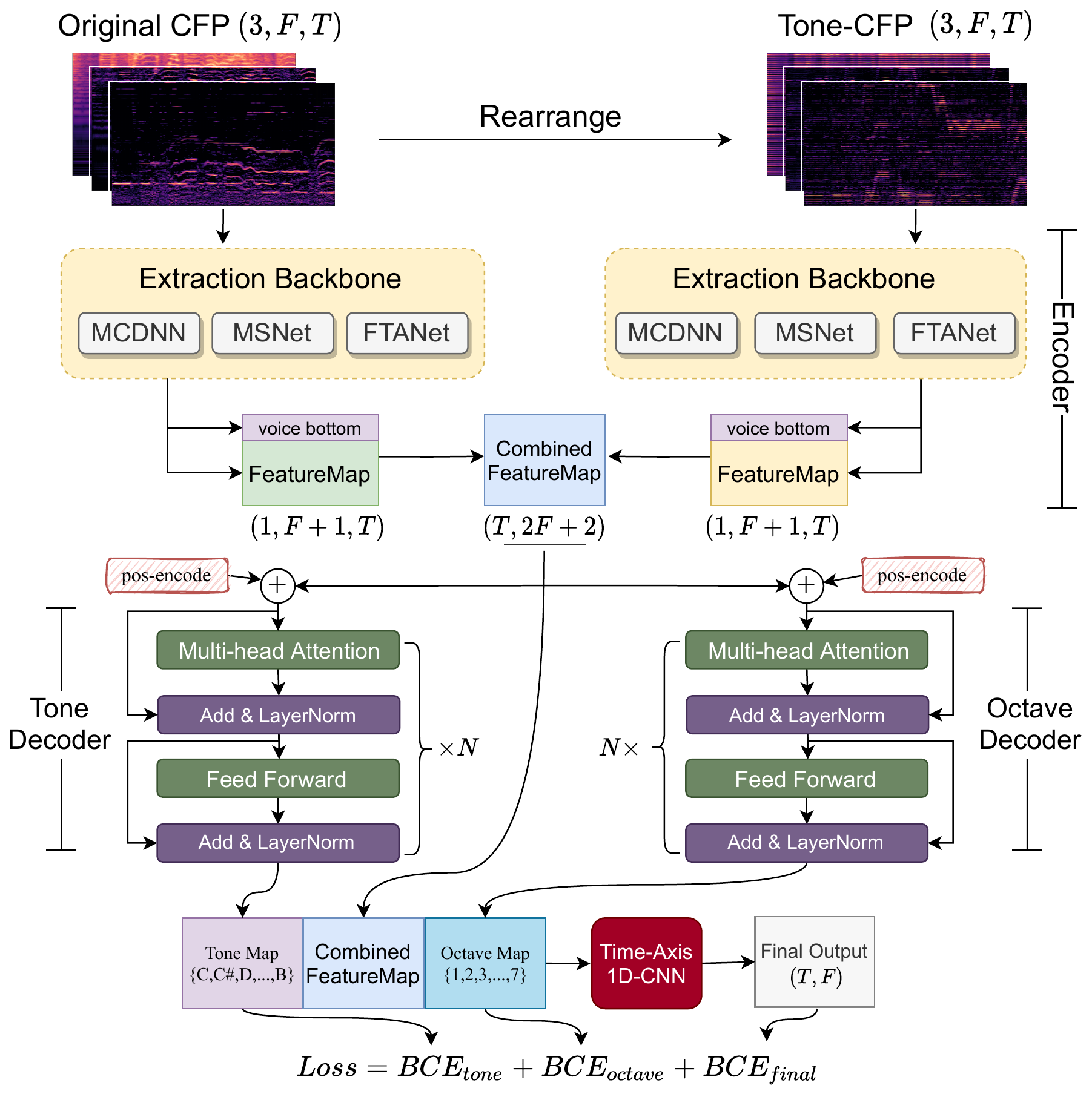}
	\caption{The architecture of TONet. Top: the CFP and TCFP; Middle: the encoder backbones and the octave-tone decoder; Bottom: the fusion mechanism and the final output.}
	\label{fig:model_arch}
\vspace{-0.5cm}
\end{figure}

As shown in the middle of Figure \ref{fig:model_arch}, the TONet encoder is a backbone model container with two parts. One part takes CFP as input, and the other part takes TCFP as input. They \textbf{do not} share the same weights and will both independently output a salient feature map of the size $(F+1,T)$, where $F$ denotes the number of frequency bins, $T$ denotes the number of time frames. An extra one-dimensional feature is added to serve as an independent voiced bottom (or non-melody core) \cite{hsieh2018streamlined, GaoYC20, ftanet}. Then, We concatenate both feature maps from CFP and TCFP as the combined feature map $(T,2F+2)$ after the axis permutation. It is passed through the tone decoder and the octave decoder for the next predictive step. Figure \ref{fig:model_arch} shows three possible backbones for the TONet encoder: MCDNN \cite{kum2016melody}, MSNet \cite{hsieh2018streamlined}, and FTANet \cite{ftanet}. One advantage of the TONet is that its encoder is not limited to one specific model, which provides flexibility for people to use it with their own models. In the experiment, we will use these three different backbone models to train different TONets and evaluate their improvements. 

\vspace{-0.3cm}

\subsection{The TONet Decoder and Tone-Octave Fusion}
As shown in the bottom of Figure \ref{fig:model_arch}, the TONet decoder passes the combined salient feature $(T, 2F+2)$ into two branches: tone decoder and octave decoder. For each branch, we employ the transformer-encoder structure with self-attention modules \cite{vaswani2017attention} to capture the temporal dependency of the salient feature. It contains a multi-head self-attention layer, a feed-forward layer, and two normalization layers with residual shortcuts \cite{resnet}. The absolute positional encoding is embedded with the input as an addition vector to denote the position information of input tokens. Multiple self-attention modules can be connected back and forth for $N$ times. 
For the hyper-parameters of the transformer, we set the contextual dimensional size $d_k=1024$ and the multi-head size $h = 8$ in each self-attention layer, where each head vector is of $\frac{1024}{8}=128$ dimensions. We set the number of modules $N=2$. After we obtain the contextual representation output $C$, we use fully-connected layers to decode the tone/octave into presence probability maps. 

The decoder outputs the predictive maps of tones and octaves, respectively, with the same time frame length $T$. We concatenate the tone map, the octave map, and the previous combined feature map as $(2F+2+P+O, T)$ after permutation, where $P,O$ are the numbers of tones and octaves. As depicted in Figure \ref{fig:model_arch}, we use a time-axis 1D-CNN layer to process this gathered map into the final output $(F,T)$. The target of this procedure is to integrate the tone and octave information into the final output. The model receives more information of tones and octaves and changes the predictions, acting as an information fusion mechanism. We do not singly use the prediction of tones and octaves without the original salient map because \cite{hsieh2018streamlined} has proved CFP brings a huge improvement on the performance with its high resolution in the frequency axis. The final loss function is the binary cross entropy with three related variables:
\begin{align}
\begin{split}
  L = &BCE(Y^{'}_{tone}, Y_{tone}) + BCE(Y^{'}_{octave}, Y_{octave}) \\
  &+ BCE(Y^{'}_{final}, Y_{final})
\end{split}
\end{align}
where $Y^{'}$ is the predictive probability of tone/octave/final map, $Y$ is the ground truth. The tone/octave ground truth can be obtained mathematically by the 12-Tone Equal Temperament \cite{12tone} with the standard pitch \textit{A4}=440 Hz.


\begin{table}
\resizebox{1.01\columnwidth}{!}{

\begin{tabular}{c|cccc|cccc}
\hline 

\hline
Datasets & \multicolumn{4}{c|}{ADC 2004} & \multicolumn{4}{c}{MIREX 05}  \\
\cline{3-4} \cline{7-8}  
Metrics &  RPA & RCA & ROA$^*$ & OA & RPA & RCA & ROA$^*$ & OA  \\
\cline{1-2} \cline{3-9} 
MCDNN &  61.6 & 63.1 & 62.6 & 66.4 & 64.1 & 64.4 & 65.2 & 75.4 \\
MCDNN$^{D}$ & 62.6 & 63.8 & 63.8 & 68.2 & 65.6 & 66.1 & 67.1 & 76.2 \\
MCDNN$^{TC}$ & 62.9 & 63.6 & 63.6 & 67.5 & 65.1 &  65.5 &  66.6 & 75.7  \\
MCDNN$^{F}$  & 69.9 & 73.2 & 78.8 & 70.8 & \textbf{74.1} & \textbf{74.6} & \textbf{78.4} & 78.8 \\ 
\textbf{TO-MCDNN} & \textbf{71.5} & \textbf{73.7} & \textbf{81.7} & \textbf{72.8} & 73.1 & 73.3 &  77.7 & \textbf{78.9} \\
\hline
MSNet & 76.4 & 76.9 & 82.6 & 79.0  & 76.7 & 76.9 & 81.4 & 81.7 \\
MSNet$^{D}$ & 75.8 & 76.6 & 82.4 & 78.0 & 78.0 & 78.2 & 83.1 & 82.6 \\
MSNet$^{TC}$ & 77.3 & 78.1 & 85.4 & 78.3 & 79.1 & 79.4 & 83.6 & 83.1 \\
MSNet$^{F}$ & 80.8 & 81.1 & \textbf{90.2} & 81.2 & 81.0 & 81.1 & 86.8 &  84.3 \\
\textbf{TO-MSNet} & \textbf{82.6} & \textbf{82.9} & 89.4 & \textbf{82.6} & \textbf{82.5} & \textbf{82.8} &  \textbf{88.1} & \textbf{85.3} \\
\hline
FTANet & 76.3 & 76.5 & 81.6 & 77.4  & 77.8 & 77.8 & 82.5 &    84.4  \\
FTANet$^{D}$ & 74.7 & 74.8 & 82.1 & 78.2 &  79.5 & 79.6 & 83.6 & 84.3\\
FTANet$^{TC}$  & 76.3 & 76.4 & 80.4 & 78.8 & 80.0 & 80.2 &  84.9 & 84.7 \\
FTANet$^{F}$ & 78.5 & 78.6 & \textbf{86.1} & 80.2 & 83.4 & 83.6 &  89.0 & 84.9 \\
\textbf{TO-FTANet} & \textbf{80.1} & \textbf{80.4} & \textbf{86.1} & \textbf{82.3} & \textbf{83.8} &  \textbf{84.0}  &  \textbf{89.7} &  \textbf{86.6} \\
\hline 

\hline

\end{tabular}}
\caption{Results of Ablation Study on ADC2004 and MIREX05 test sets.}
\label{tab:aba_test}
\vspace{-0.5cm}
\end{table}

\vspace{-0.3cm}
\section{Experiments}
\subsection{Experimental Setup}
We choose three backbone models for the encoder: MCDNN \cite{kum2016melody}, MSNet \cite{hsieh2018streamlined}, and FTANet \cite{ftanet}. There are two reasons to choose these three models: (1) they have compatible input and output formats for our encoder, which is a prerequisite of using TONet; And (2) their network structures vary from simple to complex, thus allowing us to explore the TONet's improvements in different network complexities.

For the training, we choose all 1,000 Chinese karaoke clips from the MIR-1K\footnote{https://sites.google.com/site/unvoicedsoundseparation/mir-1k} and 35 vocal tracks from the MedleyDB \cite{bittner2014medleydb}. For the testing, we select tracks that contains human singing melodies from ADC2004, MIREX05\footnote{https://labrosa.ee.columbia.edu/projects/melody} and a non-overlapping subset of MedleyDB. As a result, 12 clips in ADC2004, 9 clips in MIREX05, and 12 clips in MedleyDB are selected as test sets.

For hyper-parameters of CFP, we use 8,000 Hz sampling rate, 768-sample window size, and 80-sample hop size to compute the basic STFT. The duration of each segment is 1.28 secs ($T$=128). Then we select the number of CFP frequency bins $F$=360, with 60 bins per octave and 6 octaves in total. The start and stop frequencies are 32.5 Hz and 2,050 Hz (from \textit{C1} to \textit{B6}). The number of tone classes is 13 with 12 tones (C, C\#, ..., B) and one non-melody class ($P$=13). The number of octave classes is 7 with 6 octaves and one non-melody class ($O$=7). 

For hyper-parameters of the backbone models, we adjust some parameters to fit our settings. For MCDNN, we change the input feature from spectrogram to CFP/TCFP to make it consistent with MSNet and FTANet. In MSNet and FTANet, we set three max-pool/max-unpool kernels as ($4 \times 1$), ($3 \times 1$), ($6 \times 1$). The setting of the TONet decoder is mentioned in section 2.3. The kernel size of the 1D-CNN before the TONet's final output is $5$ with the padding $2$.
 
Our model is trained and tested in NVIDIA RTX 2080Ti GPUs. All models are implemented in Pytorch. For trainable parameters update, the Adam optimizer \cite{kingma2014adam} is used with a learning rate of 0.0001 and a batch size of 16. We use the following metrics for performance evaluation: overall accuracy (OA), raw pitch accuracy (RPA), raw chroma accuracy (RCA), voicing recall (VR) and voicing false alarm (VFA) from \texttt{mir\_eval} library \cite{raffel2014mir_eval}. In the literature \cite{salamon2014melody}, OA is often considered more important than other metrics.  

\vspace{-0.4cm}
\subsection{Ablation Study}
\vspace{-0.2cm}
As shown in Table \ref{tab:aba_test}, we conduct four ablations with different superscripts by following a paradigm on verifying each component's effectiveness. First, since the TONet uses two backbones in the encoder, we firstly only use the TONet's encoder and feed two same CFPs into it to make the parameter size equal (as $D$). Then, we replace one CFP input with TCFP but still without the decoder to test only the effectiveness of TCFP (as $TC$). And we add the tone-octave decoder but remove the TCFP encoder branch to test only the effectiveness of tone-octave fusion (as $F$). Finally, the $TO\text{-}$ model is the full TONet with the corresponding backbone.

Due to the page limit, we select ADC2004 and MIREX05 as the datasets for ablation studies, and the MedleyDB test set is evaluated in the following experiment. We only shows the RPA, RCA and OA as the most relevant metrics for ablation studies. Additionally, since we leverage the TONet to reduce the octave errors, we provide another new metric called \textbf{raw octave accuracy (ROA)} to measure \textbf{only the octave accuracy} of each prediction regardless of tone errors. Note that similar to RCA, unvoiced frames will also not be calculated in ROA.

\begin{table}[t]
\centering
\resizebox{0.8 \columnwidth}{!}{
\begin{tabular}{c|ccccc}
\hline 

\hline
Dataset & \multicolumn{5}{c}{ADC 2004} \\
\cline{2-6} 
Metrics &  VR  & VFA$\downarrow$ & RPA & RCA & OA \\
\hline
MCDNN \cite{kum2016melody} & 65.0 & 10.5 & 61.6 & 63.1 & 66.4  \\
MSNet \cite{hsieh2018streamlined} & 84.7 & \textbf{9.6} & 76.4 & 76.9 & 79.0 \\
FTANet \cite{ftanet} & 83.2 & 13.3 & 76.3 & 76.5 & 77.4 \\
DSM \cite{bittner2017deep} & 89.2 & 51.3 & 75.4 & 77.6 & 69.8 \\
\textbf{TONet} & \textbf{91.8} & 17.1 & \textbf{82.6} & \textbf{82.9} & \textbf{82.6}\\
\hline 

\hline
Dataset & \multicolumn{5}{c}{MIREX 05} \\
\cline{2-6} 
Metrics &  VR  & VFA$\downarrow$ & RPA & RCA & OA \\
\hline
MCDNN \cite{kum2016melody} & 66.5 & \textbf{4.6} & 64.1 & 64.4 & 75.4\\
MSNet \cite{hsieh2018streamlined} & 83.6 & 9.4 & 76.7 & 76.9 & 81.7 \\
FTANet \cite{ftanet} & 83.9 & 5.2 & 77.8 & 77.8 & 84.4 \\
DSM \cite{bittner2017deep} & 91.4 & 45.3 & 75.7 & 77.0 & 68.4\\
\textbf{TONet} & \textbf{91.6} & 8.5 & \textbf{83.8} &  \textbf{84.0}  &  \textbf{86.6} \\
\hline 

\hline
Dataset & \multicolumn{5}{c}{MEDLEY DB} \\
\cline{2-6} 
Metrics &  VR  & VFA$\downarrow$ & RPA & RCA & OA \\
\hline
MCDNN \cite{kum2016melody} & 37.4 & \textbf{5.3} & 34.2 & 35.3 & 62.3 \\
MSNet \cite{hsieh2018streamlined} & 53.8 & 12.4 & 47.2 & 48.4 & 66.9\\
FTANet \cite{ftanet} & 60.4 & 10.1 & 54.4 & 55.2 & 71.4 \\
DSM \cite{bittner2017deep} & \textbf{86.6} & 44.3 & \textbf{70.2} & \textbf{72.4} & 64.8  \\
\textbf{TONet} & 64.2 &  12.5 & 56.6 & 58.0 & \textbf{71.6} \\
\hline 

\hline
\end{tabular}}
\caption {The comprehensive results of TONet and benchmark models on three datasets (test sets). }  
\label{tab:exp_result}
\vspace{-0.5cm}
\end{table}


From Table \ref{tab:aba_test}, we make three observations. First, we compare $TC$-ablation with $D$-ablation models in MSNet and FTANet. Most RCAs have increased by 1.5-3\%, indicating that the structure of grouping harmonics in TCFP helps reduce the tone error. However, in MCDNN, TCFP hardly brings any improvement. The reason behind it is that MCDNN \textbf{does not} use the CNN layer, thus there is no local receptive field. TCFP and CFP make no difference to MCDNN. Secondly, we compare $F$-ablation with the original models, all OAs in both test sets have increased a lot by 2-4\%. All ROAs and RCAs have increased significantly by 5-16\%. This shows that the tone-octave fusion mechanism greatly improves the tone and octave accuracy. Thirdly, among all models, the full TONets in three models has the best OA, ROA and RCA except one ROA drop in TO-MSNet on ADC2004. On the basis of the tone-octave fusion, TCFP can further improve the performance because the prediction of tone is further improved by the use of TCFP. However, we also notice that when only introducing TCFP, the OA does not show improvement even though the RCA is improved. One possible reason is that the TCFP may lose some octave information without the tone-octave fusion, which reduces the final OA. This emphasizes the combination of TCFP and the tone-octave decoder to introduce better performance. 

\begin{figure}
    \centering
	\begin{subfigure}[b]{0.23 \textwidth}
		\includegraphics[scale=0.27]{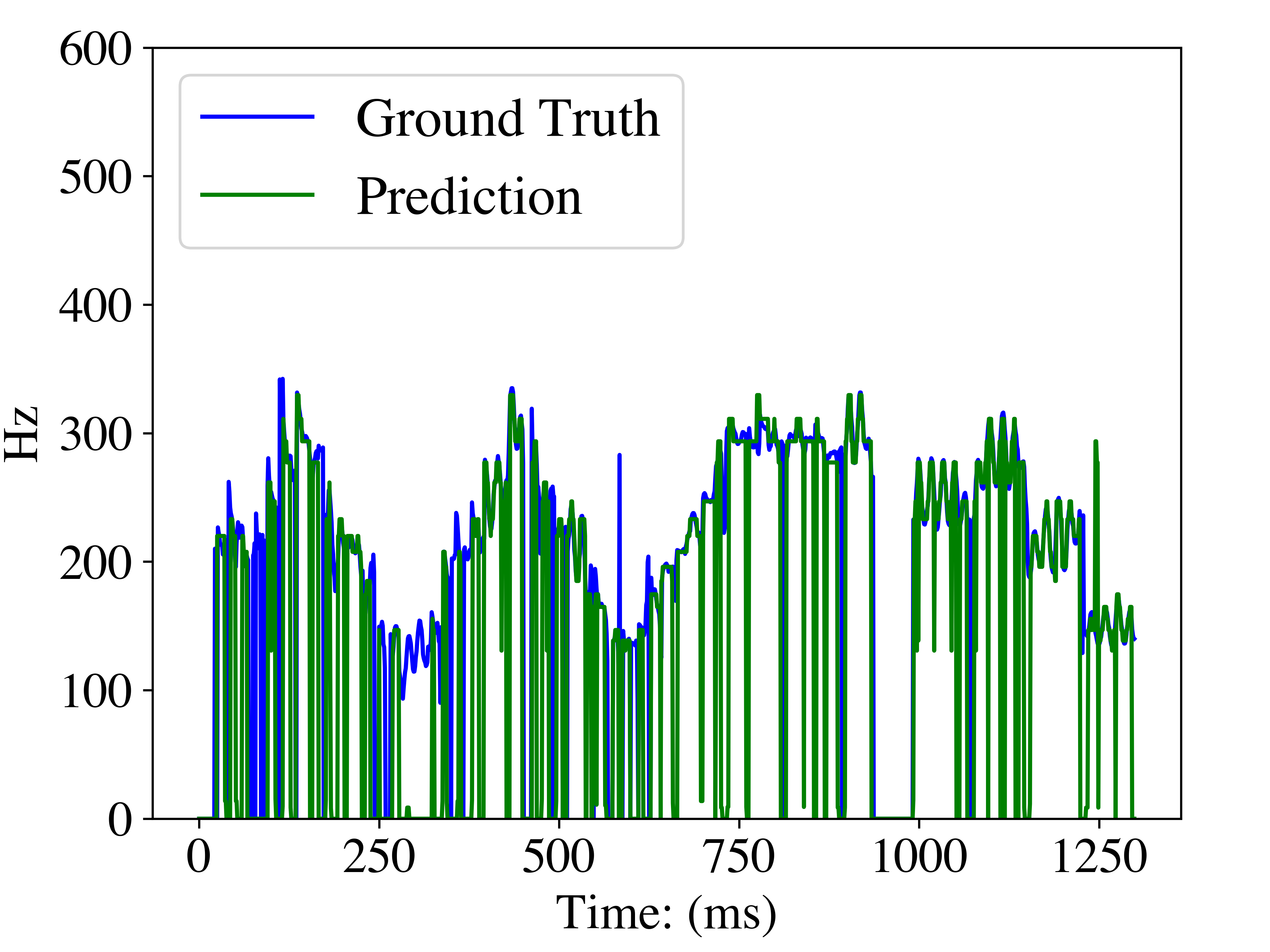}
		\caption{Opera\_male3 by TO-MSNet.}
	\end{subfigure}
	\begin{subfigure}[b]{0.23 \textwidth}
		\includegraphics[scale=0.27]{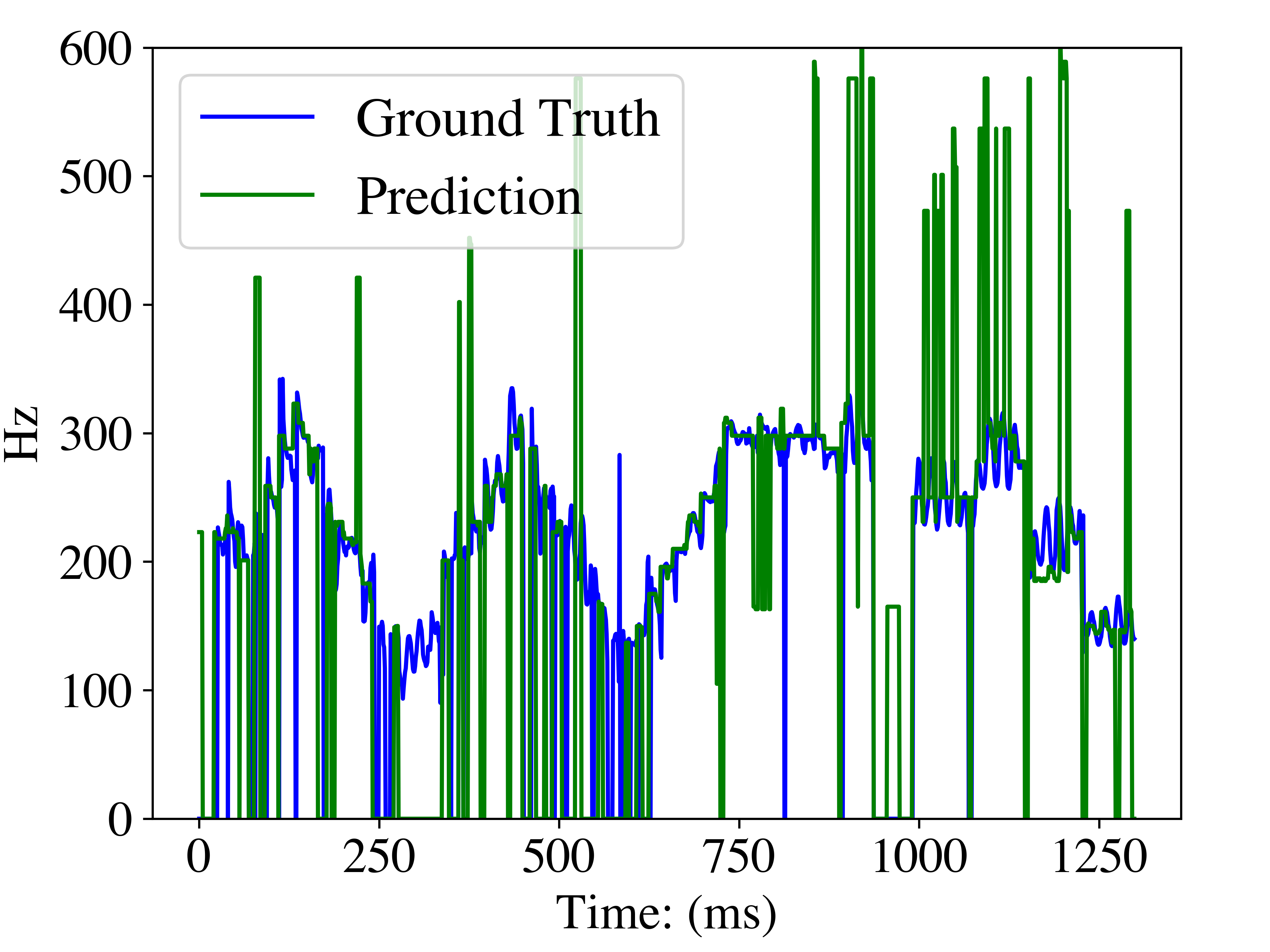}
		\caption{Opera\_male3 by MSNet \cite{hsieh2018streamlined}}
	\end{subfigure}
	\begin{subfigure}[b]{0.23 \textwidth}
		\includegraphics[scale=0.27]{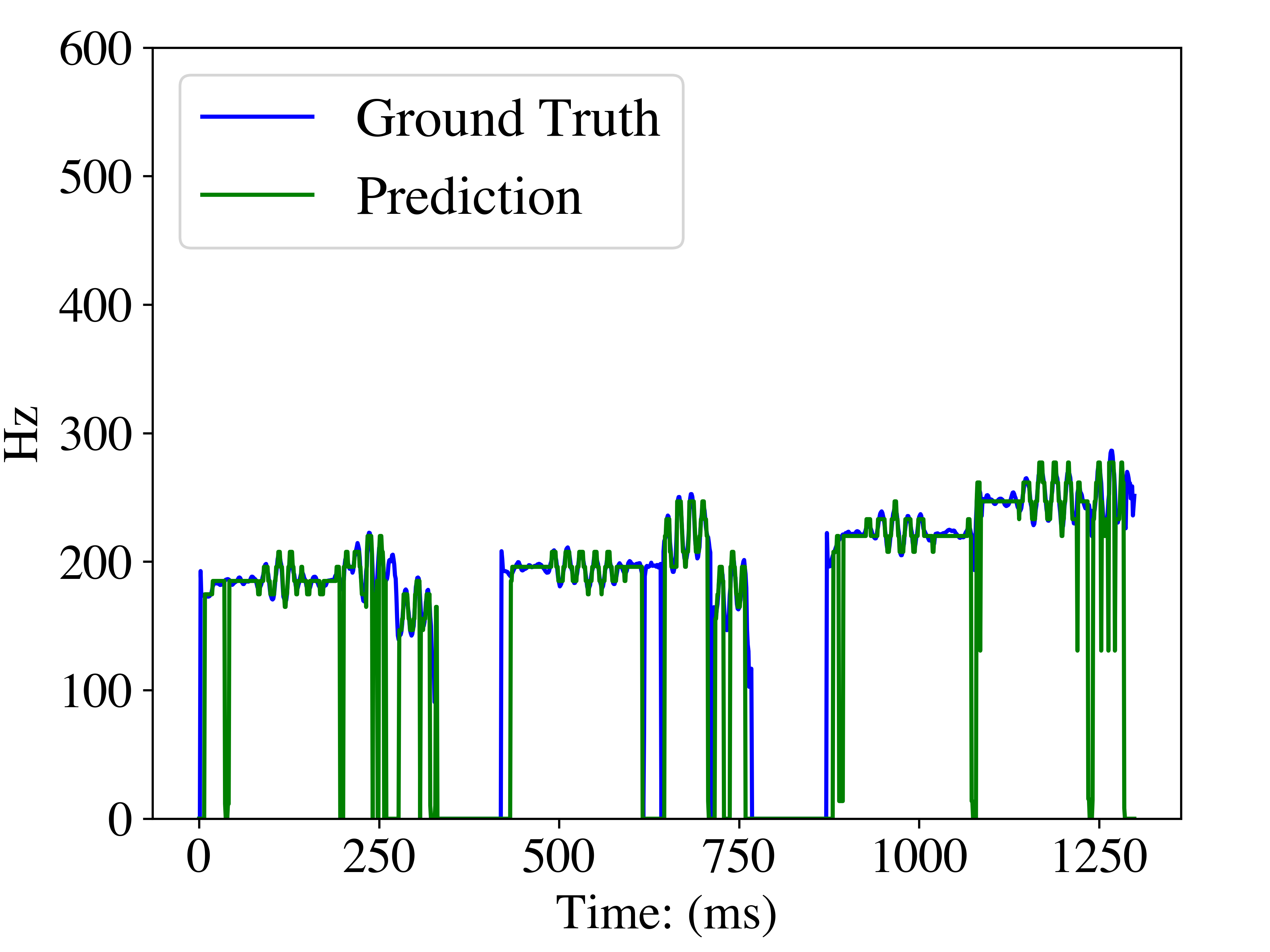}
		\caption{Opera\_male5 by TO-MSNet.}
	\end{subfigure}
	\begin{subfigure}[b]{0.23 \textwidth}
		\includegraphics[scale=0.27]{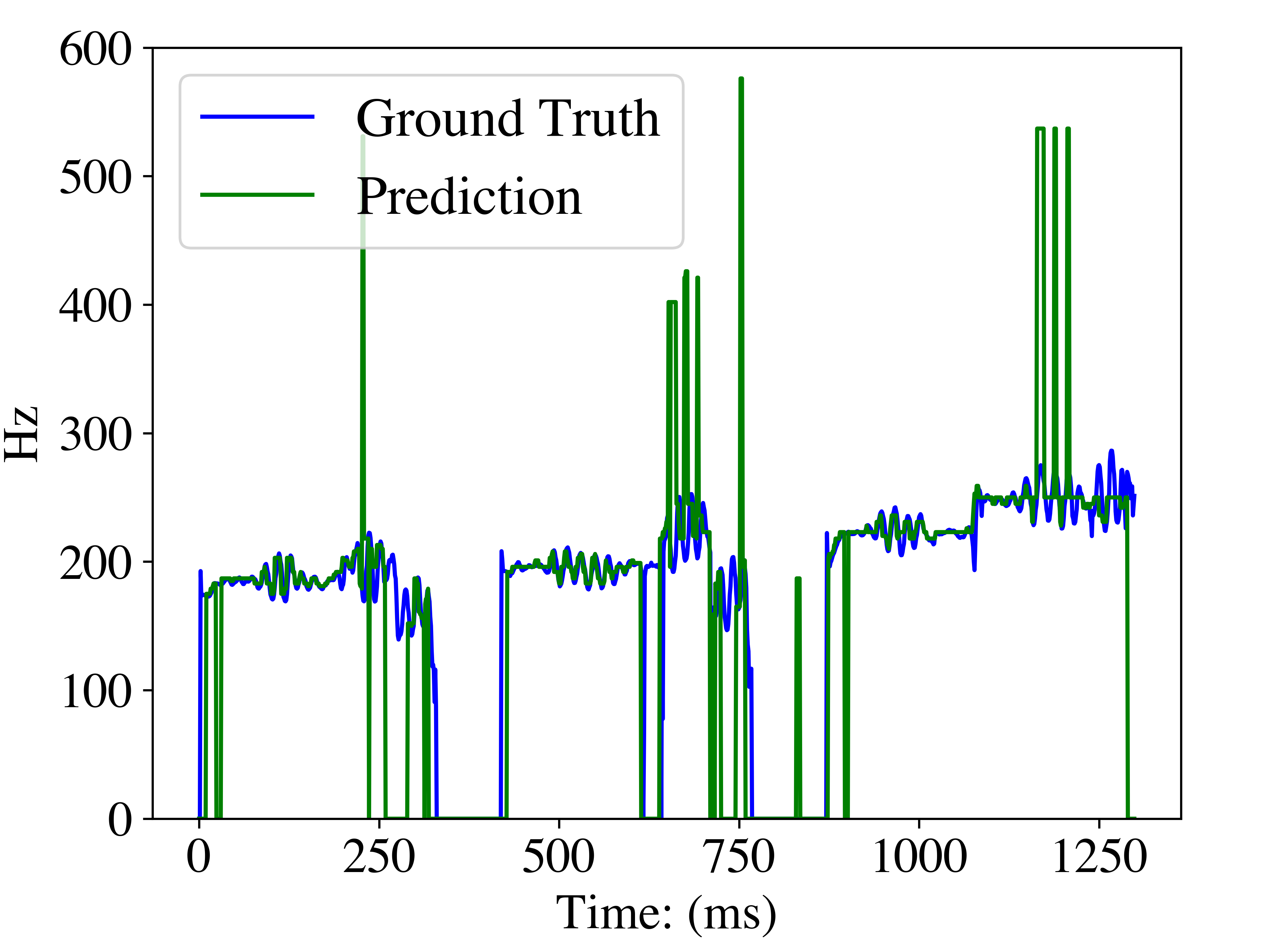}
		\caption{Opera\_male5 by MSNet \cite{hsieh2018streamlined}}
	\end{subfigure}
	
	\caption{Visualization of singing melody extraction results on two opera songs using TO-MSNet and MSNet.}
	\label{visual_plot}
	\vspace{-0.5cm}
\end{figure}

\vspace{-0.3cm}
\subsection{Comprehensive Performance Comparison}

The clearer results with five standard metrics among TONets and benchmark models are listed in Table \ref{tab:exp_result}. We select the best model among TO-MCDNN, TO-FTANet, and TO-MSNet as the final TONet. We also compared TONet with the Deep Salience Model (DSM) \cite{bittner2017deep}, which also adopts a harmonic related representation HCQT to improve its performance. Our proposed TONets achieves the best performances on ADC2004 and MIREX05 except VFA. However, in the MedleyDB test set, our model has lower VR, RPA and RCA than DSM. On the one hand, it shows the stronger ability of HCQT in DSM to capture the harmonic and related features in the MedleyDB songs. On the other hand, we find that most CFP-based melody extractors \cite{hsieh2018streamlined,ftanet} do not generalize well in the MedleyDB as with low RPA and RCA. We leave this discovery of different audio features in TONet as a future work.

An intuitive visualization on the task improvements is shown in Figure \ref{visual_plot}. We plot the predictive frequencies over the time and the ground truths by TO-MSNet and MSNet on two opera songs: ``opera\_male3.wav'' and ``opera\_male5.wav'' in ADC2004. When focusing on the octave errors (i.e., vertical jumps in the contours), we can observe that there are fewer octave errors in graph (a) and (c) produced by TO-MSNet than in graph (b) and (d) produced by MSNet. Moreover, between 1000-1250ms in diagram (d), we can find errors that incorrectly predict the frequency bin near the correct one, which are correctly predicted in diagram (c). 

Through above results, we can conclude that the performance gains of TONet can be largely attributed to the reductions of tone errors and octave errors by introducing TCFP and self-attention tone-octave fusion.


\vspace{-0.3cm}
\section{Conclusion}

In this paper, we proposed a novel Tone-Octave network that divides the traditional musical note into tone and octave representations and utilizes them to improve the singing melody extraction results. With our proposed tone-octave fusion mechanism and Tone-CFP, tone errors and octave errors are reduced and self-attention modules has indeed improved the performance of the model. Experimental results of three different backbones show that the TONet achieves promising results on three public datasets. Furthermore, TONet is proposed as a plug-and-play framework for researchers to test and optimize additional models. In the future work, we will focus on validating TONet on more datasets and devising more solid audio features (e.g., music/audio embedding in other tasks \cite{ke-sep, ke-cmg}) to be used in TONet.

\bibliographystyle{IEEEbib}
\bibliography{refs}

\begin{thebibliography}{10}

\bibitem{ke-recom}
Ke~Chen, Beici Liang, Xiaoshuan Ma, and Minwei Gu,
\newblock ``Learning audio embeddings with user listening data for
  content-based music recommendation,''
\newblock in {\em {ICASSP} 2021}.

\bibitem{ke-cover2}
Xingjian Du*, Ke~Chen*, Zijie Wang, Bilei Zhu, and Zejun Ma,
\newblock ``Bytecover2: Towards dimensionality reduction of latent embedding
  for efficient cover song identification,''
\newblock in {\em {ICASSP} 2022}.

\bibitem{ke-thesis}
Ke~Chen,
\newblock ``Controllable monophonic music generation via latent variable
  disentanglement,''
\newblock {\em Master Thesis Archive}, 2021.

\bibitem{ke-sketchnet}
Ke~Chen, Cheng{-}i Wang, Taylor Berg{-}Kirkpatrick, and Shlomo Dubnov,
\newblock ``Music sketchnet: Controllable music generation via factorized
  representations of pitch and rhythm,''
\newblock in {\em {ISMIR} 2020}.

\bibitem{TabrikianDD04}
Joseph Tabrikian, Shlomo Dubnov, and Yulya Dickalov,
\newblock ``Maximum a-posteriori probability pitch tracking in noisy
  environments using harmonic model,''
\newblock {\em {IEEE} Trans. Speech Audio Process. 2004}.

\bibitem{kum2016melody}
Sangeun Kum, Changheun Oh, and Juhan Nam,
\newblock ``Melody extraction on vocal segments using multi-column deep neural
  networks.,''
\newblock in {\em {ISMIR} 2016}.

\bibitem{lu2018vocal}
Wei-Tsung Lu and Li~Su,
\newblock ``Vocal melody extraction with semantic segmentation and
  audio-symbolic domain transfer learning,''
\newblock in {\em {ISMIR} 2018}.

\bibitem{su2018vocal}
Li~Su,
\newblock ``Vocal melody extraction using patch-based cnn,''
\newblock in {\em {ICASSP} 2018}.

\bibitem{chen2019cnn}
Ming-Tso Chen, Bo-Jun Li, and Tai-Shih Chi,
\newblock ``Cnn based two-stage multi-resolution end-to-end model for singing
  melody extraction,''
\newblock in {\em {ICASSP} 2019}.

\bibitem{hsieh2018streamlined}
Tsung-Han Hsieh, Li~Su, and Yi-Hsuan Yang,
\newblock ``A streamlined encoder/decoder architecture for melody extraction,''
\newblock in {\em {ICASSP} 2019}.

\bibitem{jdc-melody}
Sangeun Kum and Juhan Nam,
\newblock ``Joint detection and classification of singing voice melody using
  convolutional recurrent neural networks,''
\newblock {\em Applied Sciences}, vol. 9, no. 7, 2019.

\bibitem{GaoYC20}
Ping Gao, Cheng{-}You You, and Tai{-}Shih Chi,
\newblock ``A multi-dilation and multi-resolution fully convolutional network
  for singing melody extraction,''
\newblock in {\em {ICASSP} 2020}.

\bibitem{ftanet}
Shuai Yu, Xiaoheng Sun, Yi~Yu, and Wei Li,
\newblock ``Frequency-temporal attention network for singing melody
  extraction,''
\newblock in {\em {ICASSP} 2021}.

\bibitem{bittner2017deep}
Rachel~M Bittner, Brian McFee, Justin Salamon, Peter Li, and Juan~Pablo Bello,
\newblock ``Deep salience representations for f0 estimation in polyphonic
  music.,''
\newblock in {\em {ISMIR} 2017}.

\bibitem{tone-octave-percp}
Valter Prpic, Mauro Murgia, Matteo~De Tommaso, Giulia Boschetti, Alessandra
  Galmonte, and Tiziano Agostini,
\newblock ``Octave bias in pitch perception: The influence of pitch height on
  pitch class identification,''
\newblock {\em Perception}, vol. 45, no. 9, pp. 1060--1069, 2016.

\bibitem{vaswani2017attention}
Ashish Vaswani, Noam Shazeer, Niki Parmar, Jakob Uszkoreit, and Llion et~al.
  Jones,
\newblock ``Attention is all you need,''
\newblock in {\em {NeurIPS} 2017}.

\bibitem{chromafeature}
Roger~N Shepard,
\newblock ``Circularity in judgments of relative pitch,''
\newblock {\em Journal of the Acoustical Society of America}, vol. 36, no. 212,
  pp. 2346--2353, 1964.

\bibitem{resnet}
Kaiming He, Xiangyu Zhang, Shaoqing Ren, and Jian Sun,
\newblock ``Deep residual learning for image recognition,''
\newblock in {\em 2016 {IEEE} Conference on Computer Vision and Pattern
  Recognition, {CVPR} 2016}. 2016, pp. 770--778, {IEEE} Computer Society.

\bibitem{12tone}
H.~Floris Cohen,
\newblock ``Simon stevin's equal division of the octave,''
\newblock {\em Annals of Science}, vol. 44, no. 5, pp. 471--488, 1987.

\bibitem{bittner2014medleydb}
Rachel~M Bittner, Justin Salamon, Mike Tierney, Matthias Mauch, Chris Cannam,
  and Juan~Pablo Bello,
\newblock ``Medleydb: A multitrack dataset for annotation-intensive mir
  research.,''
\newblock in {\em {ISMIR} 2014}.

\bibitem{kingma2014adam}
Diederik~P Kingma and Jimmy Ba,
\newblock ``Adam: A method for stochastic optimization,''
\newblock {\em {ICLR} 2014}.

\bibitem{raffel2014mir_eval}
Colin Raffel, Brian McFee, Eric~J Humphrey, Justin Salamon, Oriol Nieto, Dawen
  Liang, Daniel~PW Ellis, and C~Colin Raffel,
\newblock ``mir\_eval: A transparent implementation of common mir metrics,''
\newblock in {\em {ISMIR} 2014}.

\bibitem{salamon2014melody}
Justin Salamon, Emilia G{\'o}mez, Daniel~PW Ellis, and Ga{\"e}l Richard,
\newblock ``Melody extraction from polyphonic music signals: Approaches,
  applications, and challenges,''
\newblock {\em IEEE Signal Processing Magazine}, vol. 31, no. 2, pp. 118--134,
  2014.

\bibitem{ke-sep}
Ke~Chen, Xingjian Du, Bilei Zhu, Zejun Ma, Taylor Berg{-}Kirkpatrick, and
  Shlomo Dubnov,
\newblock ``Zero-shot audio source separation through query-based learning from
  weakly-labeled data,''
\newblock in {\em {AAAI} 2022}.

\bibitem{ke-cmg}
Ke~Chen, Gus Xia, and Shlomo Dubnov,
\newblock ``Continuous melody generation via disentangled short-term
  representations and structural conditions,''
\newblock in {\em {IEEE} {ICSC} 2020}.

\end{thebibliography}

\end{document}